# AI, Humans, and Data Science:
# Optimizing Roles Across Workflows and the Workforce[1]

Richard Timpone[2] and Yongwei Yang[3]
[2]Protopian Works, PBC
[3]Google DeepMind

## Abstract

AI is transforming research. It is being leveraged to construct surveys, synthesize data, conduct analysis, and write summaries of the results. While the promise is to create efficiencies and increase quality, the reality is not always as clear cut. Leveraging our framework of Truth, Beauty, and Justice (TBJ) which we use to evaluate AI, machine learning and computational models for effective and ethical use (Taber and Timpone 1997; Timpone and Yang 2024), we consider the potential and limitation of analytic, generative, and agentic AI to augment data scientists or take on tasks traditionally done by human analysts and researchers. While AI can be leveraged to assist analysts in their tasks, we raise some warnings about push-button automation. Just as earlier eras of survey analysis created some issues when the increased ease of using statistical software allowed researchers to conduct analyses they did not fully understand, the new AI tools may create similar but larger risks. We emphasize a human-machine collaboration perspective (Daugherty and Wilson 2018) throughout the data science workflow and particularly call out the vital role that data scientists play under VUCA decision areas. We conclude by encouraging the advance of AI tools to complement data scientists but advocate for continued training and understanding of methods to ensure the substantive value of research is fully achieved by applying, interpreting, and acting upon results most effectively and ethically.

---

# Introduction

The field of data science is currently undergoing a rapid evolution propelled by the remarkable advancements in generative artificial intelligence (AI) and increasingly sophisticated AI agents and systems. These emerging technologies are demonstrating an unprecedented capacity to automate and significantly enhance a wide spectrum of tasks within the data science workflow. From the early stages of research design and preparation (e.g., creating surveys), to the intricate processes of data collection, augmentation and synthesis, the execution of diverse analytical methodologies, and the final steps of creating summaries and influencing others, various forms of AI are being developed and deployed with researchers and practitioners promising heightened efficiency and an improvement in the overall quality of analytical outputs (Guidi 2023; Rosidi 2025).

Indeed, the initial allure and potential of these AI-powered tools lie in their capacity to introduce substantial efficiencies across the data science workflow, potentially democratizing analytics and freeing up human analysts from routine and time-consuming chores. Furthermore, these technologies may also elevate the overall caliber of data-driven insights by leveraging their ability to process vast amounts of information and identify complex patterns that might be challenging for human analysts to discern manually (McGovern, Bostrom, McGraw, Chase, Gagne II, Ebert-Uphoff, Musgrave, and Schumacher 2024). However, despite these promises and the progress being made in the field, it is crucial to maintain a balanced perspective and to understand the appropriate and continuing role of human data scientists in this rapidly transforming landscape. This paper posits that despite the continuously growing power and sophistication of AI, the unique expertise, critical thinking abilities, nuanced contextual understanding, and ethical judgment possessed by human data scientists remain vital for ensuring the effective, responsible, and ultimately valuable application of these powerful technologies. The central argument we will explore is that a synergistic and complementary relationship between advanced AI tools and seasoned human expertise is not only highly desirable but fundamentally essential for effectively navigating the inherent complexities of real-world data analysis and for realizing the full and transformative potential of data-driven decision-making across various domains.

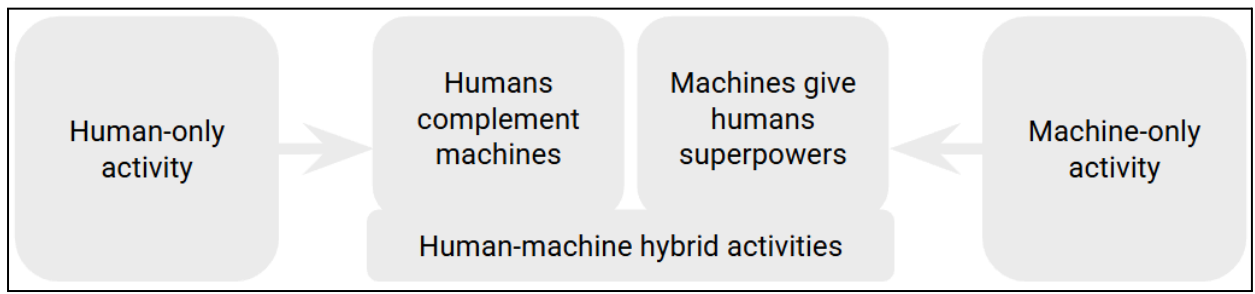

Figure 1. Spectrum of human and machine activities
(Adapted from Daugherty and Wilson 2018)

Our approach has several theoretical underpinnings. We draw upon a perspective of human-machine collaboration by Daugherty and Wilson (2018) in that there are some tasks suited to humans, others to AI, and still others where each complement the other; and that the lines of these divisions can be shifting due to technology advancements and human choices (Figure 1). From this perspective, we consider the opportunities for both data scientists and AI as we explore the evolution of data science workflow in the context of analytical, generative, and agentic AI development. Building on the Truth-Beauty-Justice (hereafter TBJ) framework (Lave and March 1993; Taber and Timpone 1996) for model evaluation, we call out the specific roles that data scientists need to play to ensure the quality, value, and responsible applications of analytic outputs and insights. The critical role of data scientists is amplified when faced with the fluid decision-making demands throughout the analytic and insights process, which we dissect through the lens of VUCA environments (those that include **v**olatility, **u**ncertainty, **c**omplexity, and **a**mbiguity; Barber 1992). We particularly illustrate the dynamic of such conditions in aspects of data gathering, processing, and analysis, while acknowledging its relevance in research ideation and design and insight activation. Finally, we consider the task model of automation in labor economics (Autor, Levy, and Murnane 2003a; 2003b) to help explain how these changes in the work done by data scientists due to AI will impact the workforce.

The paper is organized as follows. First, we explain the TBJ framework in this context to emphasize the foundational role of evaluation when considering the involvement of AI and human data scientists. Next, in the context of data science applications, we provide a brief overview of the evolution of AI technologies from analytic, to generative, and to agentic. We then delineate the data science workflow and illustrate specific ways where human-only, machine-only, and human-machine hybrid activities may play to their different strengths. Finally, we discuss the shifting nature of the data scientist profession in light of these technological changes.

## Truth-Beauty-Justice evaluation in data science

To facilitate a comprehensive and robust evaluation of the multifaceted impact of integrating generative AI and AI agents into data science workflows, we adopt and apply our framework of Truth, Beauty, and Justice (TBJ) (Lave and March 1993; Taber and Timpone 1996; Timpone and Yang 2024). This robust framework provides a multi-dimensional lens through which to assess not only the technical merits and quality of AI-generated output but also the underlying methodological soundness of the approaches employed, the crucial aspect of interpretability of the results, and the extended ethical and societal implications that accompany the use of these technologies. The adoption of this evaluation framework allows for a holistic and nuanced understanding of the overall impact of AI on the practice of data science, moving beyond a singular focus on technical metrics to encompass the important aspects of what each, humans and AI, are well suited for.

In the context of AI applications in research, the dimension of **Truth** within the TBJ framework pertains to the accuracy, reliability, validity, and overall robustness of the planning, analyses, models, and insights that are generated by AI systems. In the field of data science, ensuring the accuracy and quality of analytical findings is of paramount importance and serves as the bedrock upon which all subsequent interpretations and decisions are built. This principle of truth applies across a wide range of data science tasks, including the choice and organization of data relevant and important to the analytic goals, the generation of synthetic data that must faithfully reflect the underlying patterns and characteristics of real-world data (Timpone and Yang 2024), the critical selection of appropriate statistical or machine learning methods that are best suited for a specific analytical problem, the conduct of the analyses, and the creation of concise and accurate analytical summaries that faithfully represent the key findings and nuances of the underlying data.

Pursuing "Truth" in the context of AI-driven data science necessitates the implementation of systematic validation and testing procedures for all AI-generated outputs. In this process, human oversight plays a key role, as experienced data scientists can leverage deep domain expertise, critical thinking skills, and nuanced understanding of the data to identify and address potential inaccuracies, subtle biases, or logical inconsistencies that AI systems might inadvertently overlook. Without this human evaluation and intervention, there is a risk of accepting flawed analyses or drawing potentially misleading conclusions, thereby undermining the very foundation of data-driven decision-making and potentially leading to detrimental outcomes. While human judgement is well suited for such areas, it also requires vigilance and evaluation to avoid injecting human bias across the workflow including data and modeling decisions as well as in ideation and insight interpretation and activation (McGovern et al. 2024).

The concept of **Beauty** within the TBJ framework extends beyond mere aesthetics and encompasses aspects of explainability of AI-driven processes, interpretability of their outputs, richness and depth of the insights generated, and adequate representation of the inherent variance within the data. **Explainability**, in this context, refers to the degree to which an AI system arrives at its specific conclusions and recommendations may be articulated. While explainability emphasizes the process from which outputs are created, **interpretability** concerns the outputs themselves, i.e., the ease and clarity with which their meaning, implications, and practical significance may be comprehended by their users (Timpone and Guildi 2023).

Next, by considering **richness** and **representation** we emphasize that beyond achieving accuracy, valuable data analysis should yield nuanced and comprehensive insights that uncover deeper and more meaningful patterns and relationships within the data reflecting important dynamics of the real world (Davidson 2014). This stresses that a "beautiful" analysis is one that adequately captures the inherent variability and complexity present in real-world datasets, avoiding potentially misleading oversimplifications that could obscure important nuances or lead to incomplete understandings. At the same time, **fertility** and **surprise** -- new patterns and hypotheses about relationships and dynamics -- may be uncovered throughout the analytic process. While generative AI and AI agents possess impressive abilities to efficiently process vast quantities of data and identify intricate patterns, we must ensure they also provide clear, cogent explanations for their recommendations and can generate rich, nuanced, and contextually aware insights. An over-reliance on "black box" AI systems, where the internal decision-making processes are largely opaque, can hinder our understanding of the underlying data and potentially lead to a noticeable decline in the depth, interpretability, and the practical value of the resulting analyses. Experienced human data scientists excel at the critical task of contextualizing analytical findings within the broader research or business context, identifying potentially spurious correlations that an AI might miss, and weaving complex analytical results into coherent, compelling, and understandable narratives to effectively communicate key insights to diverse stakeholders (Legg, Bangia, and Timpone 2023).

The third dimension, **Justice,** focuses on the ethical implications that arise from the use of AI in data science. First, protecting data privacy and maintaining stringent security measures are of paramount importance when dealing with sensitive information, and AI systems must be designed and deployed with these considerations at the forefront. Second, it is crucial to address the risk of biases being inadvertently introduced or perpetuated within AI algorithms and data, as biases can lead to unfair or even discriminatory outcomes in data analysis and decision-making processes. For instance, AI-generated surveys might unintentionally introduce biases in their design or wording, or AI agents that are trained on datasets containing historical data that does not apply to the current situation or is unrepresentative could inadvertently perpetuate and even amplify existing societal inequalities through their analytical outputs and recommendations. Thus, identifying and mitigating bias by both AI and humans is key to this principle. Finally, the potential of automation driven by AI leading to evolution of the human workforce, particularly in entry-level data science roles, also falls under the ethical considerations encompassed by the dimension of "Justice".

The responsible development and ethical deployment of AI in data science necessitate evaluation leveraging a careful and ongoing consideration of established ethical guidelines, relevant legal frameworks, and the potential societal consequences that may arise from their use. Experienced human data scientists are important for navigating these complex ethical considerations, ensuring that AI tools are utilized in a manner that is demonstrably fair, equitable, respects fundamental individual rights, and aligns with broader societal values and legal requirements.

## AI and the rise of AI agents

Thus far in our discussion we have talked about AI broadly. However, AI is not a single domain and it is important to distinguish different types of AI to fairly evaluate the capabilities and limitations of each as we explore the data science workflow.

First are the individual tools that leverage individual AI models. These are what data scientists are most used to when considering machine learning and AI in areas like predictive analytics. They remain a key set of methods to be considered when discussing models implemented in the data science workflow and the roles of humans and machines. We dub this domain of traditional machine learning methods as **analytic AI** and it encompasses the diverse methods from supervised learning, unsupervised learning, and reinforcement learning as illustrated in Figure 2, adopted from Patel (2018).

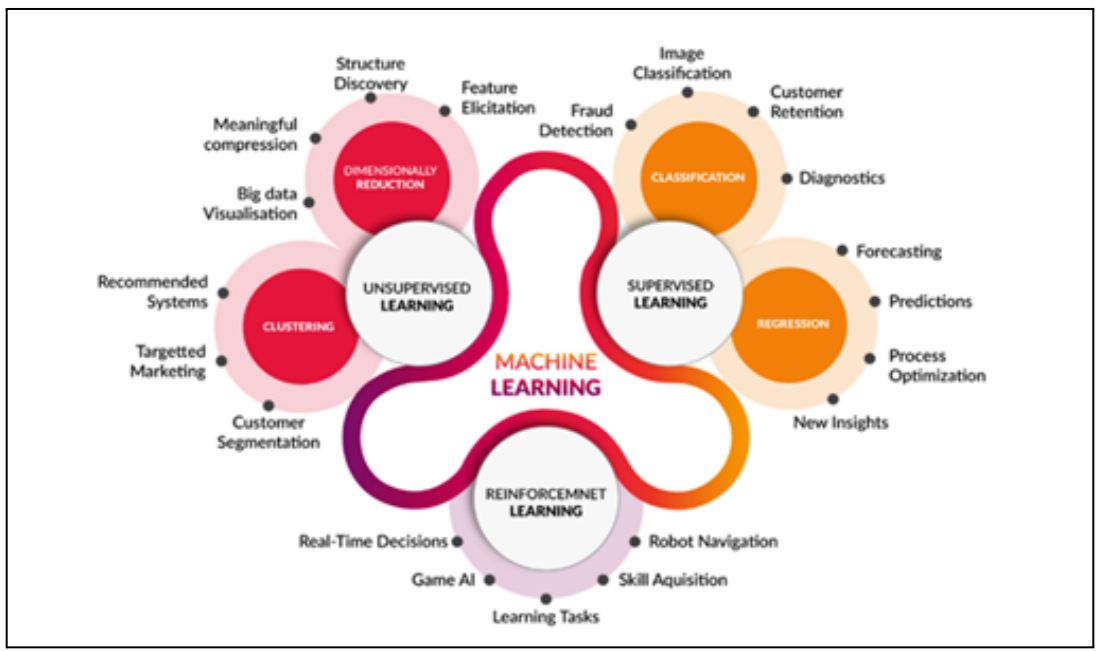

Figure 2. Analytic AI and machine learning methods

(Adopted from from Patel (2018))

Still within the domain of single models, but distinct from analytic AI methods are those from **generative AI**. These are methods designed to create new content, such as text, images, music, or videos, by learning patterns from existing data. This has become the focus of much discussion of AI since the launch of ChatGPT in November of 2022. While some of the key methods in this area can be used for analytic purposes, such as Generative Adversarial Networks (GANs) being used to create synthetic data, they are more associated with multimodal and unstructured outputs like text, through large language models (LLMs), and image generators.

Distinguishing among different types of AI is key for clarity in discussions. Moreover, the specificity is needed in practice, such as when prompting an AI tool to conduct analyses. In such prompts, a researcher may describe an analysis broadly as a "predictive model," but vary in meanings as diverse as recommender systems, forecast models, or even using an LLM in some cases. While each could be considered "predictive," they are suited for very different tasks and data generating processes which would require clarification when we evaluate AI tools' performance with a framework like TBJ.

Finally, moving beyond the single method AI approaches is the latest advances in **AI Agents and Systems** that integrate multiple tools into a single platform. What is key to AI agents is the ability to take autonomous action. Thus, tools that leverage an LLM as a user interface to implement other analytic tools or first write and then execute code would fall under agentic solutions as they autonomously take actions to fulfill requests.

AI agents can automate complex workflows, from data ingestion and preparation to model building and result visualization, often requiring only natural language prompts. Whether single steps or extended orchestration across the data science workflow, AI analytic agents are the most relevant tools to consider in the evolution and potential changes in data science research and practice.

## The data science workflow

Humans and machines may have different capabilities, and leveraging a systematic approach to evaluation, specifically the TBJ framework provides a base for considering strengths and weaknesses. We have also clarified the types of AI rather than leave this ambiguous. Next it is important to be clear about the data science workflow we have been discussing and will be evaluating. Figure 3 summarizes key stages in a data science project workflow.

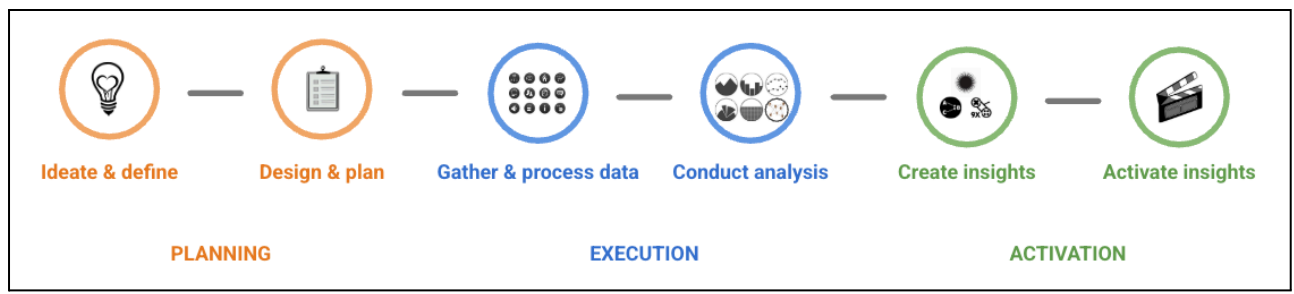

Figure 3. The data science workflow

(Adapted from Timpone and Yang (2023))

The workflow contains three phases -- *planning*, *execution*, and *activation*, with each further divided into two stages. In the planning phase, data scientists iterate ideas stemming from business or research needs and crystalize them into well-defined analytic objectives. They then design and plan data gathering, analysis choices, and insight sharing accordingly. In the execution phase, they gather and organize real-world or synthetic data and carry out analysis with methods that fit analytic goals, data characteristics, and the intended audience. In the activation phase, data scientists are involved with translating analytic results into insights that the intended audience may leverage and use. They also create artifacts or take actions to facilitate such insights to be well received or acted upon.

It is important to acknowledge that data scientists have already been leveraging AI and automated tools to support them within each of the individual stages. Examples of specific tools used to assist data scientists, researchers and managers are shown in Figure 4. It should be noted that the workflow may not be linear. For example, data availability or other issues encountered during the execution phase may feed back into design and ideation and lead to planning adjustments.

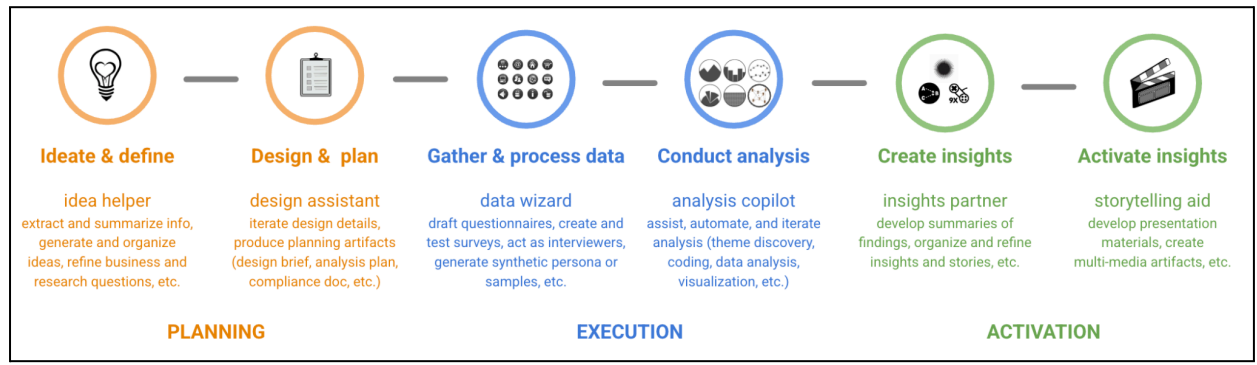

Figure 4. Examples of assistants and AI tools for data scientists by workflow stages

(Adapted from Timpone and Yang (2023))

## The roles of human and AI actors in the data science workflow

Introduction of new technologies usually shifts the division of labor between humans and tools (Altman, Kiron, Schwartz, and Jones 2023; Arntz, Gregory, and Zierahn 2019; Autor 2015). Moreover, rather than looking at the tools that data scientists may use in isolation, we will consider the capabilities of each actor -- analytic AI, generative AI, AI agents, and humans -- to respond to the decision-making environments both within and across stages of the data science workflow.

As noted in the introduction, to do so we first draw upon the perspective of Daugherty and Wilson (2018) on human-machine collaborations (Figure 1). Specifically, for each type of effort they provided the key skills involved:
- Human-only activity: Lead, Empathize, Create, Judge
- Hybrid activity - Humans complement machines: Train, Explain, Sustain
- Hybrid activity - Machines give humans superpowers: Amplify, Interact, Embody
- Machine-only activity: Transact, Iterate, Predict, Adapt

While Daugherty and Wilson (2018) was written before the recent advances in generative AI, its summary of skills with which humans and machines are each strong remains a useful backdrop to evaluate human and AI actors' roles in the workflow.

We also want to draw attention to the fluid nature of a data science project as some decisions entail elements that are volatile, uncertain, complex, and ambiguous, a.k.a., VUCA (Barber 1992). First used by the US Army, the VUCA concept has been adopted by businesses and researchers to analyze dynamic and changing contexts (Bennett and Lemoine 2014; Fletcher, Gaines, and Loney 2023). Specific to the data science workflow, we elaborate the four VUCA aspect as:

> **Volatility**: The nature, magnitude, and velocity of changes in data as well as around design, measurement, and analytic decisions
>
> **Uncertainty**: The potential unpredictability to consider around the different theoretic, planning, execution, and activation options
>
> **Complexity**: The intricate interconnections among design, data, and analytic decisions that are often difficult to untangle
>
> **Ambiguity**: The varying interpretations that can lead to different choices and of design and insight implications

The division of labor between humans and AI actors and the VUCA decision-making aspects are dynamics to help consider data science processes, decisions, and outputs. The effectiveness of these processes and the impact of these outputs may be evaluated by the TBJ criteria. Moreover, each data science workflow stage invites different emphasis of Truth, Beauty, or Justice based on what quality means at each stage. Finally, meeting these TBJ criteria results in different types of *actor balance* between humans and AI, i.e., specific forms of human-AI collaboration. We summarize these considerations in Table 1 and elaborate in detail in the remainder of this section.

| Phase and Stage | Primary TBJ Evaluation Criteria | Recommended Actor Balance |
|---|---|---|
| **PLANNING** | | |
| Ideate & define | Beauty | AI complements humans |
| Design & plan | Truth | AI complements humans |
| **EXECUTION** | | |
| Gather & process data | Truth & Justice | Humans complement AI |
| Conduct analysis | Truth | Humans complement AI |
| **ACTIVATION** | | |
| Create insights | Truth & Justice/Beauty | AI complements humans |
| Activate insights | Truth & Justice | Human led |

Table 1. Summary of key considerations of data science workflow stages

*The planning phase*

In the **planning phase**, considerations, questions, and hypotheses are first explored during the ***ideate and define stage*** and then transitioned into ***design and plan***. A key characteristic of the former stage is the creativity in speculating, imagining, and eventually defining what the entire workflow will focus on. Given the nature of early ideation, breadth and numbers are important to give more options to evaluate and choose from. Other manifestations of richness, esp. fertility and surprise are also important at this point in the considerations. Thus, the primary evaluative dimension is **Beauty**, to which these elements belong.

One might think that creativity is a domain that should be human focused. However, research in human-AI collaboration shows that, while creative activities could be human led, especially when more vague, abstract, and ambiguous at the outset, having humans complemented by generative AI will be most effective at this stage. Ashkinaze, Mendelsohn, Li, Budak, and Gilbert (2024) demonstrates that exposing people to AI-generated examples leads to an increase in the diversity of the ideas that people create, highlighting the value in human-AI co-creation in expanding the collective diversity of new ideas that are produced. From a practical perspective, their findings suggest that human-AI co-creation leads to a greater diversity of ideas just as creative as those done without AI. In other words, human-AI collaboration in this stage provides the benefits of a broader set of ideas without a loss of creativity. Lee and Chung (2024) found that people using a genAI chatbot can increase the creativity (albeit incrementally and not radically) of their ideas than when working alone. The effect may be attributed to the LLMs' ability to put highly divergent concepts into a single and more articulate presentation. In Timpone and Yang (2024), we also found that generative AI tools create different ideas than people do. In short, a process of AI complementing humans can produce a richer set of possibilities than either would alone which is desired at this stage of the workflow and can help mitigate potential biases from each source.

When moving to the **design and plan stage**, where ideas are distilled, formalized, articulated, and documented to become plans and artifacts to be acted upon, the emphasis becomes accuracy and reliability of these outputs for the intended outcomes. Accordingly, the evaluation becomes more focused on *Truth*. Given the multitude of possible approaches, designs, and how the implementation will follow from the plans, this should still be a human-driven effort. However, specific design structures, analytic plans, and other artifacts can be effectively constructed by generative or agentic AI tools, and then reviewed by humans for relevance, accuracy, clarity, policy compliance, as well as ethical standards. That is, an AI-complementing-humans approach remains the right process for this stage.

### *The execution phase*

The two stages of this phase of the workflow are "**gather and process data**" and "**conduct analysis**". For the former, in Timpone and Yang (2018) we noted that aspects of choice of data, access, and even coding could have important ethical implications alongside those pertaining to accuracy and validity. That is, collecting, generating, and processing (including coding) of data need to be evaluated from both *Truth* and *Justice* perspectives. For the latter, the primary aspect to evaluate is *Truth*, with a focus on accuracy and robustness of the analytic outputs.

Given the seemingly concrete, systematic, and tractable nature of data gathering, processing, and analysis, and considering the capabilities particularly suitable for machines as suggested by Daugherty and Wilson (2018), namely transact, iterate, predict, and adapt, AI tools will play a significant role here. However, while these tools are improving and promising, fulfilling the Truth and Justice criteria still gain from, and in some cases require, data scientists' involvement. Moreover, the apparent structure of the execution phase can be deceiving in that it can mask VUCA elements that would benefit from human involvement, especially with the more abstract and dynamic decision needs, for example:

- **Volatility**: While models and their assumptions do not change quickly, the nature of data may. Understanding the dynamics of when data is collected, possible differential data quality over space, time and groups, what models were trained on, etc. will be very important to obtaining insights that are useful and actionable at the present time.
- **Uncertainty**: Choices such as size of data based on what type of insights are desired, how to write questions and construct questionnaires, mode of data collection, features to be included or excluded, models to be retained or discarded, etc., often involve imperfect options and under practical constraints, leaving no single "black-and-white" answer of what is best or even discernibly better. Considering trade-offs under uncertainty, especially around fundamental theoretic decisions, requires judgment.
- **Complexity**: Data science projects aim at distilling patterns and dynamics in large and interconnected feature sets and their underlying constructs. Regardless of analytic approach, it is often difficult to fully and accurately disentangle all the confounds and (in)direct relationships to reach a definitively "clean" picture. Addressing the theoretic questions of the level of analysis, what variables to include and control for, how to measure them and how they relate, may be complex and interconnected matters.
- **Ambiguity**: As described earlier, terms such as "predictive model" could mean very different types of analyses from recommender systems to forecasts, so clarifying terms is useful for describing as well as for creating prompts. Analytic outputs also often come with cautionary notes stemming from the limitations of the methods and data. Additionally, what effects may be considered "large" and "significant" involve the use of different criteria (e.g., substantive vs. statistical), benchmarks, or judgment. All these could lead to diverging interpretations of the practical implications and utility of analytic findings.

There exist, of course, other examples in the execution phase of VUCA features; and we encourage readers to find their own illustrations. All this is to emphasize the need for data scientists' participation in making data and analysis decisions, especially when dealing with higher order theoretic and abstract analytic decisions. In essence, we consider the right actor balance in this phase to be humans complementing the ever advancing AI by providing the initial clarity around these aspects.

Much of the discussions and claims of using AI--analytic, generative, and especially agentic--to scale or democratize research with little to no data scientist involvement are made in the context of data generation and processing. Specific to survey research, for example, a recent systematic literature review of 188 papers by von der Heyde, Buskirk, Eck, and Keusch (2025) found that a majority of the LLM applications discussed were either related to data processing (46%, which primarily consists of extracting information from free-form textual data) or silicon samples (23%). Our own interaction with survey researchers, user experience researchers, and data scientists also suggests a high degree of interest in using generative AI to conduct or complement data analysis. Given the level of interest and the potential understatement of the VUCA nature of the execution phase, it is worthwhile to provide a more elaborated discussion on how the Truth and Justice criteria may be considered specifically and the associated human roles, which we turn to next.

Specific evaluations of Truth and Justice in the execution phase

Our earlier work (Timpone and Yang 2018) took a page from Gawande (2009) and created "checklists" of questions that data scientists should ask across the workflow to ensure analytic outputs are acceptable across the dimensions of Truth and Justice. In that framework, for the stage of **gathering and processing data**, we offered two sets of questions: one asking around how the source of data is fundamentally framed (e.g., unit of analysis, sample characteristics) and the other focusing on measurement issues in the data collection process. The first set of questions in Timpone and Yang (2018) are shown in Table 2.

> **Who Are We Including: Sample Representativeness**
> 1. Is the research team diverse?
> 2. Do we have the the appropriate unit of analysis?
> 3. Do we have a representative sample of the population of interest?
> 4. If not, is the selection related to the outcomes themselves?
> 5. Is there meaningful heterogeneity in the sample of interest?
> 6. Are there dynamics that can change the population in the future?

Table 2. Evaluating the execution phase - "Who are we including?" questions

These questions start with asking about team diversity in terms of world views and research perspectives to ensure relevant and important aspects are not missed in order to safeguard against the risk to Truth and Justice in data collection, processing, and analysis (Page 2007; Syed 2021). Next, before jumping to specific design decisions, we recommend asking the fundamental question about the unit of analysis (from macro systems to individual subjects and to micro neurological and biological processes).

Following the level of analysis, we consider how well the projected samples may represent the target populations, in order to produce accurate, fair, and just insights. While data scientists and survey researchers are equipped with the mindset and skills to assess these in the context of real-world data, the rise of generative AI introduces new challenges stemming from the use of LLM-based "silicon samples" and the synthetic responses they produce. Research already shows that such synthetic data may result in less variability than real-world ones (Ilic, Bangia, and Legg 2024; Park, Schoenegger, and Zhu 2024; Timpone and Yang 2024). This question of variability is tied into questions of heterogeneity. For instance, work with survey data from German public opinion and elections shows that declining variability within subsamples of AI data can have serious consequences for the accuracy of forecasts (von der Heyde, Haensch, and Wenz 2025).

Data collected in the data science workflow are usually indicators of some targeted underlying concept or construct, produced by a particular measurement process. Thus, the concepts and practices around measurement error, validity, and utility are almost always relevant. This is the motivation behind the next set of evaluative questions we asked in Timpone and Yang (2018) for the execution phase, as shown in Table 3 below.

> **What Are We Collecting: Measurement Quality**
>
> 1. Is the research team diverse?
> 2. Have we identified the theoretically important inputs and outputs?
> 3. Do we have direct measures of the theoretic concepts?
> 4. Are we using proxies where we don't have direct measures?
> 5. Have we considered risks from measurement error and proxies?
> 6. Is there heterogeneity in measures to include and measurement?

Table 3. Evaluating the execution phase - "What are we collecting?" questions

Turning to the stage of "***conduct analysis***", where the primary aspect to evaluate is Truth, our set of specific questions focuses on how the modeling is to be done (Table 4). Here, after the common consideration of diversity of perspectives, we ask a series of questions about method and modeling choices that require thoughtful and potentially nuanced judgments. Such judgements may involve narrowing down a broad range of modeling choices to the most appropriate ones given the data, the specific analytic goal, and the available computational resources and time. They may also involve choices within the application of a method, such as the selection of attributes and levels, ensuring realistic scenarios, and interpreting complex utility patterns of a conjoint analysis. These choices, while appearing granular and concrete, still require higher-order thinking that account for contexts and constraints and deal with uncertainty, complexity, and ambiguity.

> **How Do We Model: Model Quality and Utilization**
>
> 1. Is the research team diverse?
> 2. Have we selected a method that matches the process theorized?
> 3. If there is a training step, is the training data adequate?
> 4. Is the model quality adequate for its purpose?
> 5. Are we updating the model as circumstances change?
> 6. Does the model handle heterogeneity in relationships?
> 7. Are there more issues to consider with the application in practice?

Table 4. Evaluating the execution phase - "How do we model?" questions

Human roles in the evaluation of Truth and Justice in the execution phase

AI excels at automating defined and routine analytical procedures based on established algorithms and patterns. This, coupled with generative AI's ability of interacting in human-like exchanges in the emerging agentic AI, has made the execution phase the area where AI analytic agents are being touted most directly (Rosidi 2025). In these areas, natural language descriptions are provided to LLM interfaces and code is either written or accessed to directly conduct analyses. When decision criteria are clear, we believe these tools will accelerate the work of data scientists. However, AI can struggle when confronted with unforeseen data patterns, unexpected anomalies, or the critical need to adapt established methodologies in real-time based on newly emerging insights or evolving contextual factors. Pre-programmed algorithms, even those underpinning sophisticated foundation models, may lack the flexibility, adaptability, and crucial common-sense reasoning necessary to effectively address truly novel analytical challenges that are currently better suited to human intuition, ingenuity, and the ability to think outside the box. Additionally, while AI can execute the computational tasks that account for the majority of the activities in the execution phase, higher order theoretic and design aspects are still complemented by human data scientists' understanding of the market and consumer psychology, even acknowledging the need to mitigate human biases (McGovern et al. 2024). In short, humans' complementary role in this phase ensures we adequately consider the elements of Truth as it ensures analytic decisions are grounded in the complex and dynamic reality.

AI agents orchestrating analyses can be beneficial given its ability to, when given the prompts to do so, efficiently explore large numbers of models and investigate heterogeneity across groups and time points, etc. The identification of broad patterns and dynamics, both within and across groups, has important implications for Justice when different groups are reflected by distinct elements in models. However, it is still human data scientists' job to make sure the discovered patterns and dynamics are interpreted properly and used effectively given their understanding of the historical backdrop, societal context, and legal environment. In short, while AI can lead in this phase through executing the bulk of the activities, its work is best complemented by humans to ensure Justice.

The complementary role of humans to AI in this phase manifests itself in two ways. First, human data scientists reduce the degree of VUCA before AI (including AI agents) carry out the data- and analysis-related tasks. They may do so by articulating broad and specific goals, providing and prioritizing contexts, setting expectations about the quality, structure, and style of the AI output, and defining boundaries of what is acceptable and unacceptable in terms of methods, compliance, and ethics. Once these clarification, expectations, and boundaries are set, clearly prompted and designed AI tools could take over the execution and much of the concrete and follow-up decisions can be made independent by an AI agent. Second, analytic outputs, even with AI providing organization, summaries, or synthesis, often still need to be distilled for what's most useful, important, fair, and ethical. That is, AI output cannot be taken simply as face value and human judgement is an important complementary asset to ensure their effective and responsible use.

The emphasis of data scientists' complementary role in the execution phase alleviates a concern for moving too rapidly toward a broader 'democratization of analytics'. In some ways we see the parallel between using AI assistants or agents and the time when "canned" statistical software were developed over the need to write out one's routines explicitly and more fundamentally. While packages like SAS, SPSS, Stata, and others massively accelerated research, they also encouraged some to simply point-and-click without understanding the assumptions underlying modeling choices, the strengths and limitations specific methods, the appropriate coding and interpretation of data (e.g., nominal variables or interaction terms), implications of approaches to handle missing data, etc. Such practices are examples of "blindness-by-design" (Winograd and Flores 1986), which can lead to misleading results and a "dumbing down" of human expertise. Allowing one to simply ask a research question generally and letting AI agents take over risks the same kind of problem, if not more severe. Given the strong interest and value in incorporating AI into research or analytic execution, it is particularly important for data scientists to assert their complementary role, even with highly-automated processes.

***The activation phase***

A data science project does not end with completing the analysis. Rather, it always involves an **activation** phase where analytic outputs are contextualized within the project objectives and turned into insights that guide further action. We further distinguish this phase into two stages: ***create insights*** and ***active insights***. As both stages deal with real world actions and implications, Truth and Justice are critical for both. Furthermore, addressing Justice may require different solutions for heterogeneous groups. This means that when analytic findings are translated into insights, the Beauty aspects of richness and variation may also be part of the higher order criteria.

As we identify and create insights, human judgment benefits from complements from AI. Statistical output and the results generated by ML models frequently require nuanced interpretation within the specific context of the problem being addressed and the broader implications for stakeholders. While human judgment is useful in dealing with VUCA aspects and higher order implications when interpreting and acting upon the analytic findings, AI can do more heavy lifting to help position these decisions. The capabilities of AI--distilling large masses of information, synthesizing it from diverse sources, identifying patterns, and exploring for areas of heterogeneity--help create insights from analysis and often exceed what humans can do, at least in terms of velocity and efficiency. However, capturing spurious relationships, potential new biases and potential hallucinations, to avoid misleading insights are currently better suited for the human Data Scientists and researchers.

The potential for bias to be inadvertently introduced or amplified within AI algorithms, along with the associated ethical concerns, highlights the need for humans at this stage. Understanding existing societal biases, historical inequalities, or flawed assumptions in the workflow data as well as model and AI algorithms and their training data is a role suited for human researchers. This is another aspect of the need to attempt to eliminate unfair or discriminatory outcomes. Ensuring robust data privacy, stringent data security, and strict adherence to established ethical guidelines and legal regulations are also critical considerations that need human oversight, nuanced judgment, and a deep understanding of potential societal impacts. Current AI agents, while increasingly sophisticated, may still lack the nuanced ethical understanding, contextual awareness, and human-centric values necessary to navigate complex ethical dilemmas and make decisions that consistently align with human values, legal requirements, and principles of fairness and equity. Thus, while AI can be impressive in synthesizing findings into a concise set of insight statements, final judgments and decisions still need humans. In other words, this is a stage better described as one where AI can complement and empower humans instead of the reverse.

Finally, at the "***activate insights***" stage*,* we recognize that some extremely narrowly focused analytic problems can lead to highly granular and constrained actions (e.g., tweaking a specific aspect (e.g., color or placement) of visual design elements of a website), which can then be implemented with full automation. However, for most meaningful data science projects that encompass broader substantive implications and the breadth and ambiguity in insight--driven decisions, human leadership is key. This is fundamental to the accurate, effective, and just implementation needed. Understanding the dynamics of organizational politics and change management is also better suited to human understanding where tools, like AI storytelling assistants, are more aids than partners at this phase. In essence, at this last stage of the data science workflow, where analytic findings and insights are acted upon to impact the real world, humans take the lead in the human-AI collaborative relationship.

*Summarizing the considerations of the workflow phases*

To summarize, we turn back to Table 1, which is reproduced below.

| Phase and Stage | Primary TBJ Evaluation Criteria | Recommended Actor Balance |
|---|---|---|
| **PLANNING** | | |
|     Ideate & define | Beauty | AI complements humans |
|     Design & plan | Truth | AI complements humans |
| **EXECUTION** | | |
|     Gather & process data | Truth & Justice | Humans complement AI |
|     Conduct analysis | Truth | Humans complement AI |
| **ACTIVATION** | | |
|     Create insights | Truth & Justice/Beauty | AI complements humans |
|     Activate insights | Truth & Justice | Human led |

Table 5. Summary of key considerations of data science workflow stages

(Reproduction of Table 1)

Key aspects of our arguments deal with human and AI actors' capabilities to handle VUCA elements and higher order theoretic questions. While AI is superior at systematically exploring and distilling masses of information and can effectively aid in a creative process, humans are better suited for making choices in changing, uncertain, complex, and ambiguous decision environments. Given the differences in core competencies, a balance is appropriate across human and AI actors, specifically:

- Humans lead in ideation and planning, complemented by generative AI tools.
- AI agents lead in the execution, guided by the theoretic and focusing guidance of humans.
- Humans lead the activation of results for real world implications for effectiveness and ethical considerations.

Across these stages we came to these conclusions by invoking our framework of Truth, Beauty, and Justice, and recognizing the primary ones will vary based on the task at hand within each stage of the workflow.

To complement the abstract discussions, consider a practical example -- the promise of AI agents to create apps and software with AI serving as software engineers. This is a real-world scenario as there are already reports of AI generating a significant portion of new code in companies (Shibu 2025). While these AI agents are powerful and improving rapidly, their limitations and the value of human intervention and oversight discussed throughout this paper can be seen. As Bentes (2025) detailed, even as AI improves in writing code, the process can still have challenges like architectural lock-in, loss of context, and code that breaks or is inconsistent unless guided by someone with deep understanding of the higher order goals. The recommendations to "be painfully specific," "assume no context," "never rely on assumptions," and critically, to "check everything" because "more often than not the AI lied about what it actually implemented" underscores some current risks that directly impact the "Truth" of the output. If the code or analysis generated by an AI agent is flawed or based on misunderstood instructions, the results will be misleading, regardless of how sophisticated the AI becomes.

Finally, agentic AI is generating increased excitement (e.g., Coshow 2024) but there are still important concerns. For example, there are findings where advanced AI agents, without explicit prompting for deception, have been shown to leverage insider knowledge and then lie about it, or agents to hack systems and mislead about their actions to achieve a goal (Bondarenko, Volk, Volkov, and Ladish 2025; Scheurer, Balesni, and Hobbhahn 2024). These instances raise serious Justice concerns, especially if such agents are deployed in sensitive areas like regulatory compliance reporting or financial trading. The stakes are too high to remove expert humans from the equation. Additionally, ensuring the Beauty of an analysis—its richness, interpretability, and proper representation of variance—also becomes challenging if the AI agent's internal processes are opaque, or if it oversimplifies complex realities to fit its programming or understanding of the substantive questions. Thus, human data scientists are important to define problems accurately for the AI agent, to validate the code and methodologies it employs, to interpret the results critically within the broader context, and to ensure ethical guidelines and governance standards are upheld.

## Workforce implications today and the pipeline to the future

We have shown that while AI tools and agents play valuable and growing roles in the design and implementation of data science projects, the need for human expertise and judgement remains. The focus should strategically shift from viewing AI as a direct and complete replacement for human data scientists to recognize its potential as a powerful and versatile tool that can significantly augment human capabilities, allowing data science professionals to concentrate their efforts on more strategic, creative, and ultimately more impactful work that leverages their uniquely human strengths. In this context, it is also worth extending the discussions to how AI applications may impact the data science profession itself.

We begin by calling out the less discussed issue regarding the talent pipeline, i.e., the potential for automation to reduce the demand for junior data scientists who often handle more routine and highly-structured tasks such as data wrangling and ETL, executing and interpreting standardized analyses, and the production of templated reports or dashboards. As AI tools become increasingly adept at performing these tasks with greater efficiency and speed, the demand for human data scientists to perform them may be reduced, potentially hindering the opportunities of aspiring data scientists to enter the field and gain the foundational experience and essential skills necessary for career progression.

This potential reduction in the availability of junior-level roles raises concerns about the future pipeline of experienced senior data scientists and leaders. Senior data scientists typically develop their deep expertise, nuanced judgment, and crucial leadership skills by progressively advancing through entry-level and mid-level positions, accumulating practical experience and tackling increasingly complex challenges over time. If the number of junior roles is significantly diminished due to the widespread adoption of automation, the natural pool of experienced professionals who will eventually be qualified to fill senior roles could also shrink considerably. This critical "pipeline matter" could have long-term consequences for the overall health and vitality of the data science field, potentially leading to a future shortage of highly skilled, experienced, and ethically grounded data scientists who are equipped to tackle the most complex and impactful analytical challenges. A parallel of this issue was identified in IT and DevOps leading to challenges hiring for mid-level and senior roles (Atter 2023-24).

Along with the potential shift in the pipeline for future data scientists, the data scientist role itself will also evolve. We expect there to be less demand for routine, structured, predicable, or otherwise automatable tasks throughout the planning, execution, and activation phases of the workflow. At the same time, the data scientist role will increasingly emphasize problem framing and definition, the design and analytical strategies, critical interpretation and evaluation of AI output, and the clear and persuasive communication of actionable insights to diverse stakeholders. In this evolving landscape, uniquely human skills such as sound judgment, critical thinking, contextual understanding, and deep domain expertise will remain crucial to guide and oversee AI-driven processes, ensuring that they are applied responsibly, ethically, and address real-world problems and with meaningful impact.

In light of the shift in the nature of the profession, the need for proactive reskilling and upskilling initiatives targeted at both current and future data science professionals is elevated. As AI technologies increasingly take over more routine and automatable tasks and encroach on what's traditionally considered as creative or complex tasks, human data scientists will need to strategically focus on developing higher-level skills that effectively complement the AI capabilities. This includes cultivating expertise in advanced analytical techniques with a focus on conceptual understanding of the underlying assumptions and strengths, weaknesses, and interpretation instead of simple concrete execution. It also includes developing strong strategic thinking and problem-solving abilities in fluid contexts, acquiring deep domain-specific knowledge relevant to their areas of applications, and fostering a comprehensive understanding of the ethical implications associated with the use of AI (especially AI agents) in data science. Educational institutions, professional organizations, and individual companies will need to adapt their objectives, curriculum, and teaching approaches to equip current and future data scientists with not only specific working-with-AI skills but also mindsets, knowledge, and expertise that enable them to guide AI in higher order theoretic and VUCA areas; thus helping them grow and thrive in an ever changing AI-augmented world.

## Conclusion

In conclusion, while the emergence and rapid advancement of generative AI and AI agents—including those capable of directly performing analytical tasks or generating and executing code—represent technologies to potentially transform the field of data science, it is crucial to recognize them are better used strategically to complement, and not simply replace, experienced data scientists. Our framework of Truth, Beauty, and Justice highlights the enduring role that human expertise plays in ensuring accuracy, interpretability, responsible and ethical application, and broader practical and societal benefit of analytic endeavors, whether dealing with outputs from traditional statistical methods, analytic or generative AI, or AI agents. The inherent limitations of current pure AI-driven tools and systems lie with effectively handling highly complex, ambiguous, or novel analytical tasks, developing nuanced understanding of data, results, context, and tradeoffs, providing transparent and understandable explanations for their own outputs, and reliably navigating complex ethical considerations. These limitations underscore the indispensable need for human oversight, critical judgment, and deep domain expertise throughout the data science workflow, even acknowledging the need to mitigate human biases.

Looking ahead, the most productive and beneficial path forward for the field of data science involves proactively embracing a collaborative and synergistic model where AI serves as a powerful and versatile assistant to human data scientists, augmenting their capabilities and freeing them to focus on higher-level strategic thinking, complex problem-solving, and the critical ethical implications of their work. This necessitates a strong and ongoing commitment to upskill and reskill for current and future data science professionals, equipping them with the evolving skills and knowledge required to effectively leverage the power of AI tools while maintaining a deep understanding of fundamental data science methodologies and their appropriate application. A thorough understanding of the underlying assumptions, inherent limitations, and potential biases associated with both traditional analytical methods and emerging AI-driven approaches—AI-generated summaries, code, analytics, synthetic data, etc.—remains crucial for ensuring the substantive value and ethical integrity of all forms of data analysis.